\documentclass[12pt]{article}
\usepackage{times}
\usepackage{graphicx}
\usepackage{geometry}
\usepackage{natbib}
\usepackage{aastex_hack}
\usepackage[utf8]{inputenc}
\usepackage{enumitem,amssymb}
\usepackage{ragged2e}
\newlist{thematic}{itemize}{8}
\setlist[thematic]{label=$\square$}
\usepackage{pifont}
\newcommand{\cmark}{\ding{51}}%
\newcommand{\done}{\rlap{$\square$}{\raisebox{2pt}{\large\hspace{1pt}\cmark}}%
\hspace{-2.5pt}}

\begin{document}
\raggedright
\huge
Astro2020 Science White Paper \linebreak

Cool, evolved stars: results, challenges, and promises for the next decade \linebreak
\normalsize

\noindent \textbf{Thematic Areas:} \hspace*{60pt} $\square$ Planetary Systems \hspace*{10pt} $\square$ Star and Planet Formation \hspace*{20pt}\linebreak
$\square$ Formation and Evolution of Compact Objects \hspace*{31pt} $\square$ Cosmology and Fundamental Physics \linebreak
  $\rlap{$\done$}\square$ Stars and Stellar Evolution \hspace*{1pt} $\square$ Resolved Stellar Populations and their Environments \hspace*{40pt} \linebreak
  $\square$    Galaxy Evolution   \hspace*{45pt} $\square$             Multi-Messenger Astronomy and Astrophysics \hspace*{65pt} \linebreak
  
\textbf{Principal Author:}

Name: Gioia Rau
 \linebreak						
Institution: NASA/GSFC~\&~CUA
 \linebreak
Email: gioia.rau@nasa.gov
 \linebreak
Phone: +1 (301) 286-6322
 \linebreak
 
\textbf{Co-authors:}
  \linebreak
Rodolfo Montez Jr.~(Center for Astrophysics (CfA) $|$ Harvard \&~Smithsonian), 
Kenneth Carpenter (NASA/GSFC), 
Markus Wittkowski (ESO/Garching),
Sara Bladh (Uppsala University), 
Margarita Karovska (CfA/Harvard \&~Smithsonian),
Vladimir Airapetian (NASA/GFSC), 
Tom Ayres (University of Colorado),
Martha Boyer (STScI), 
Andrea Chiavassa (OCA/Nice), 
Geoffrey Clayton (Louisiana State University), 
William Danchi (NASA/GSFC), 
Orsola De Marco (Macquarie University),
Andrea K.~Dupree (CfA/Harvard \&~Smithsonian), 
Tomasz Kaminski (CfA/Harvard \&~Smithsonian),
Joel H.~Kastner (RIT),
Franz Kerschbaum (University of Vienna),
Jeffrey Linsky (University of Colorado), 
Bruno Lopez (OCA/Nice), 
John Monnier (University of Michigan),
Miguel Montarg\`{e}s (KU Leuven),
Krister Nielsen (CUA), 
Keiichi Ohnaka (Universidad Catolica del Norte Chile), 
Sofia Ramstedt (Uppsala University),
Rachael Roettenbacher (Yale University),
Theo ten Brummelaar (CHARA/GSU), 
Claudia Paladini (ESO/Chile),
Arkaprabha Sarangi (NASA/GSFC~\&~CRESST~II-CUA), 
Gerard van Belle (Lowell Observatory), 
Paolo Ventura (INAF/OAR).
 \linebreak
 
\textbf{Abstract:}
Cool, evolved stars are the main source of chemical enrichment of the interstellar medium (ISM), and understanding their mass loss and structure offers a unique opportunity to study the cycle of matter in the Universe. Pulsation, convection, and other dynamic processes in cool evolved stars create an atmosphere where molecules and dust can form, including those necessary to the formation of life (e.g.~Carbon-bearing molecules). Understanding the structure and composition of these stars is thus vital to several aspects of stellar astrophysics, ranging from ISM studies to modeling young galaxies and to exoplanet research.\\

Recent modeling efforts and increasingly precise observations now reveal that our understanding of cool stars photospheric, chromospheric, and atmospheric structures is limited by inadequate knowledge of the dynamic and chemical processes at work. Here we outline promising scientific opportunities for the next decade that can provide essential constraints on stellar photospheres, chromospheres, and circumstellar envelopes (CSE), and tie together analyses of the spectra and interferometric and imaging observations of evolved stars.\\

\hspace{\parindent} We identify and discuss the following main opportunities:\\
(1) identify and model the physical processes that must be included in current 1D and 3D atmosphere models of cool, evolved stars;\\
(2) refine our understanding of photospheric, chromospheric, and outer atmospheric regions of cool evolved stars, their properties and parameters, through high-resolution spectroscopic observations, and interferometric observations at high angular resolution;\\
(3) include the neglected role of chromospheric activity in the mass loss process of red giant branch (RGB) and red super giant (RSG) stars and understand the role played by their magnetic fields;\\
(4) identify the important shaping mechanisms for planetary nebulae (PNe) and their relation with the parent Asymptotic Giant Branch (AGB) stars.


\pagebreak

\section{Cool, evolved stars: compelling scientific questions} 
Deciphering the structure and evolution of stellar interiors ushered the era of modern astrophysics. A new epoch in our understanding of stellar evolution lies in wait through the investigation beyond the interior into the atmosphere and circumstellar shells that form during the later evolutionary stages, specifically, the RGB, AGB, and RSG phases. During these stages the cool evolved stars swell, lose enriched stellar material, and build molecule- and dust-rich circumstellar envelopes. The resulting conditions in the photosphere, chromosphere, the outer circumstellar envelope, and their inter-connectivity, become exceedingly difficult to model theoretically and constrain observationally. However, over the past few decades, the confluence of improvements in high-resolution spectroscopic and interferometric observations and computational capabilities have brought us to the brink of several breakthroughs in our understanding of RGB, AGB, and RSG stars. Over the next decade we hope to provide answers to the following questions:

\begin{enumerate}
\item For cool K and M giant stars, is there a clear boundary where the chromosphere ends and the wind begins or do these regions overlap, and what are the terminal velocities of the winds in these stars? Do chromospheres exist in AGB stars? 
Is there a direct relationship between chromospheric activity and the presence of dust in RGB and RSG stars, and is the strength of the chromosphere effectively reduced by the presence of dust?
\textbf{We must improve our empirical understanding of photospheric, chromospheric, and outer circumstellar envelopes through systematic surveys of a large number of objects enabled by improved capacity of multiwavelength high-resolution spectroscopic and interferometric observations.}

\item What previously overlooked physical processes are essential for 1D and 3D atmospheric models of cool evolved stars to understand the mass loss process at all evolutionary stages and masses? What mechanisms drive the wind acceleration of K and M giant and supergiant stars? 
\textbf{We must establish the full gamut of essential physical processes for multi-dimensional atmosphere modeling, including the oft-neglected role of magnetic fields and chromospheric activity in their atmosphere and mass-loss processes.}

\item What are the effects of companions on mass-loss processes for cool evolved stars? What mechanism(s) could shape AGB stars into PNe?  How do the shaping processes within the circumstellar envelopes (CSE) of RSGs influence their subsequent core-collapse supernova explosions? Which are the progenitors of PNe and Type~Ia SNe?
\textbf{We must develop strong connections and constraints on the mass loss mechanisms of these stars to later times of  evolution, such as the shaping of planetary nebulae, novae, core-collapse and Type~Ia supernovae, and subsequent populations of stars and planets.} 
\end{enumerate}

\section{Empirical challenges to our understanding} 
\begin{figure}
\begin{center}
\includegraphics[angle=0, clip=true, width=0.55\hsize]{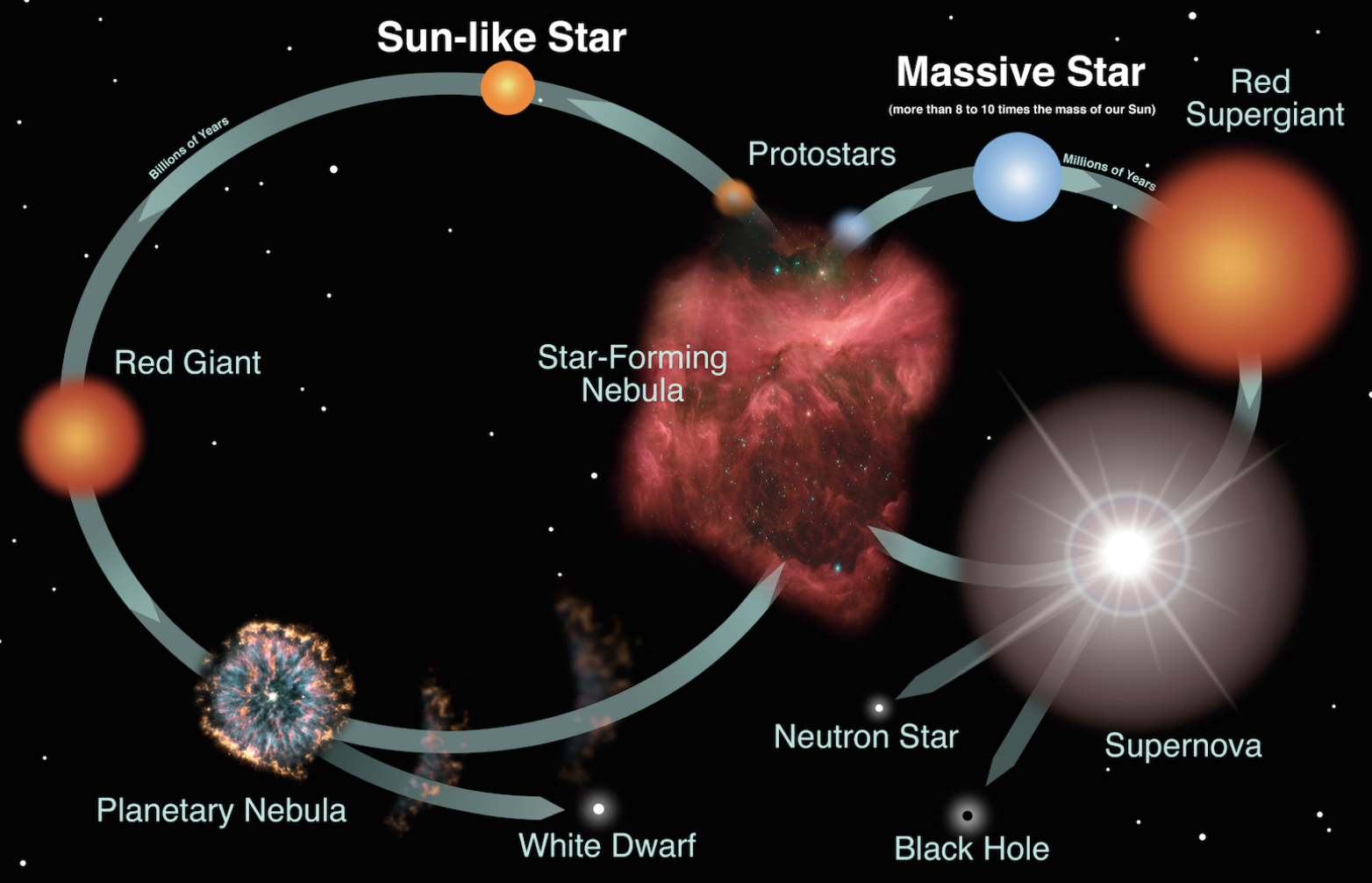}
\caption{\label{cycle}The life cycle of gas and dust in the Universe, and stages in the life of a single star. (Figure credits: NASA/JPL, Astronomical Society of the Pacific).}
\end{center}
\end{figure}

Cool, evolved stars are major contributors to the chemical enrichment of the Universe. They supply the ISM with heavy chemical elements, molecules, and dust through mass-loss provided by their stellar winds. These chemical elements are essential components for the cycle of matter in the Universe (see Fig.~\ref{cycle}).

\hspace{\parindent} AGB, RGB, and RSG stars are located in the Hertzsprung-Russell-Diagram at cool effective temperatures between $\sim~2500$~K and $4500$~K along Hayashi tracks, have large extended radii up to hundreds $R_{\odot}$, and, depending on their initial mass, cover a large range of luminosities. Due to the low temperatures of AGB and RSG stars, molecules and dust can form in their atmospheres where stellar winds can expel this material into the interstellar medium. The mass-loss rates range from $4\cdot10^{-8}$ to $8\cdot10^{-5}~$M$\odot$~yr$^{-1}$ for AGB stars \citep{ramstedtolofsson}, and from $2\cdot10^{-7}$ to  $3\cdot10^{-4}~$M$\odot$~yr$^{-1}$ for RSGs \citep{debeck10}.

\begin{figure}
\begin{center}
\includegraphics[angle=0,  width=0.45\hsize]{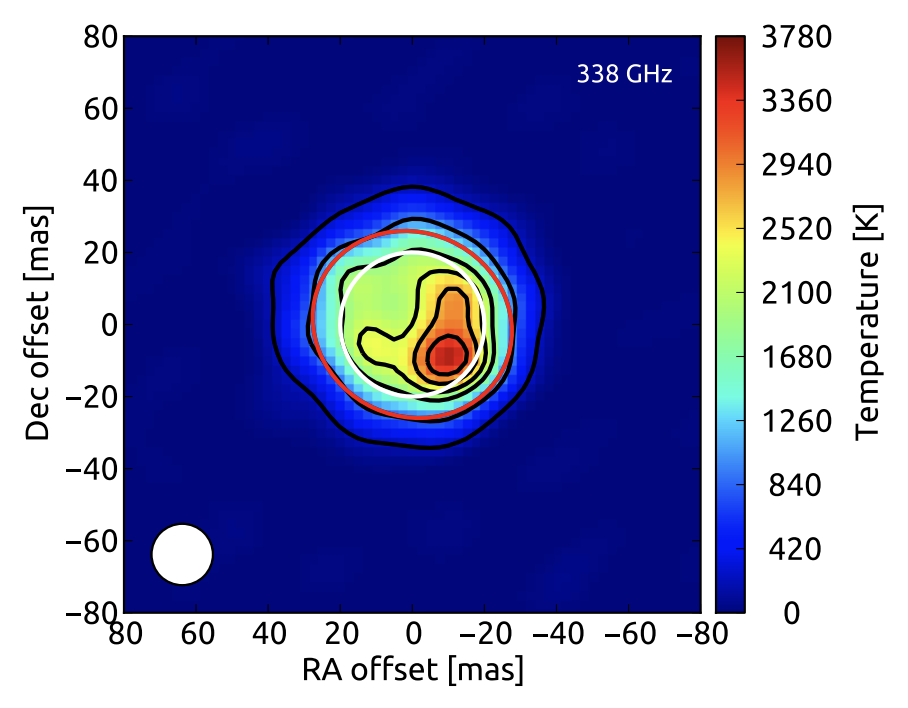}
\includegraphics[angle=0,  width=0.35\hsize]{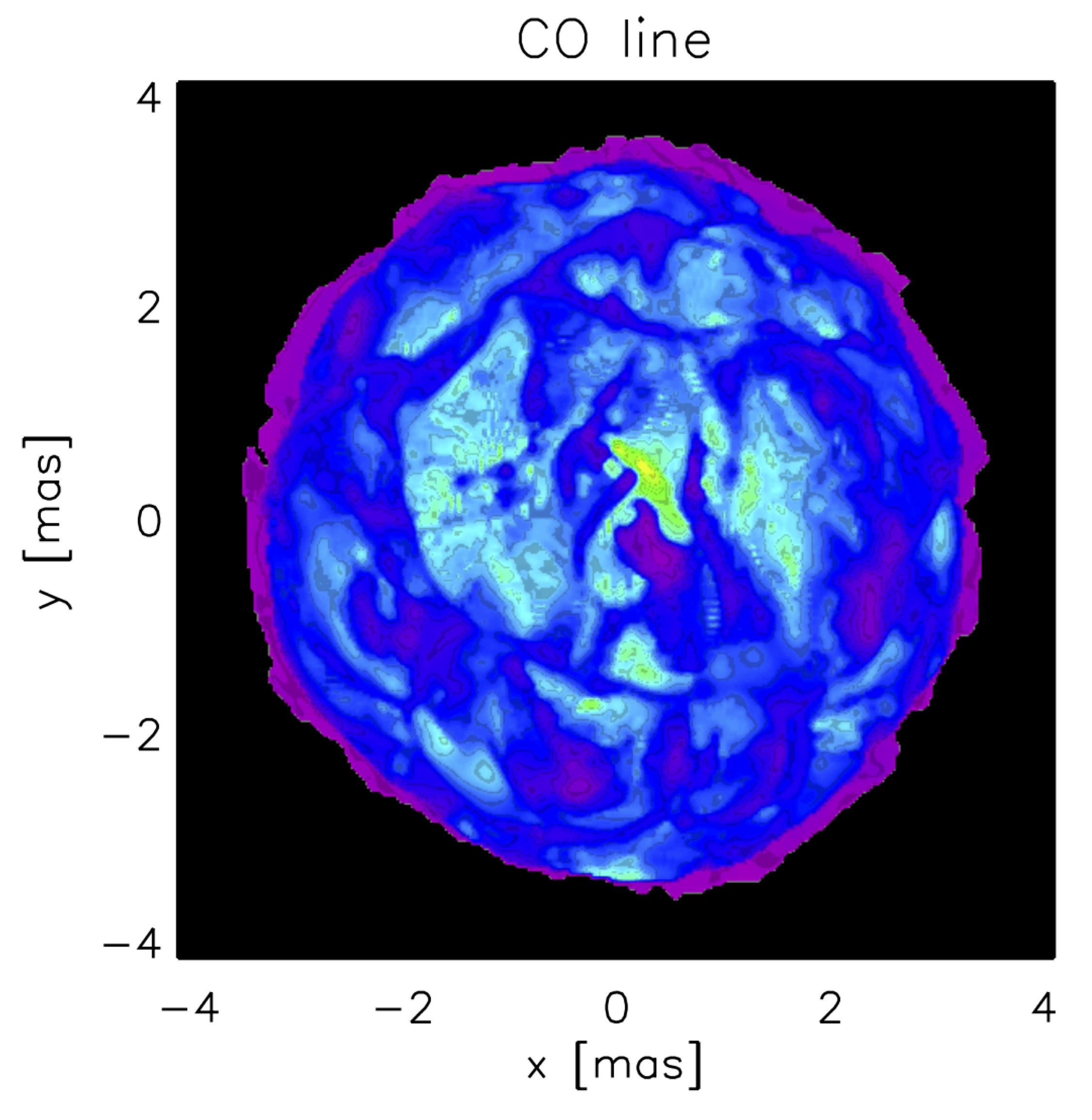}
\caption{\label{COBOLD_2}{\it Left:} Brightness temperature distribution on the surface of the AGB star W Hya from \citep{2017NatAs...1..848V} and based on interferometric ALMA observations. {\it Right:} 3D radiative-hydrodynamical (RHD) simulations of convection for one snapshot of a COBOLD model from \citealp{arroyo-torres15}. The image shows the surface intensity as seen at the CO (2-0) line at $2.294~\mu$m.}
\end{center}
\end{figure}

\hspace{\parindent} AGB, RGB, and RSG stars are affected by pulsation and convection. Most AGB stars are pulsating with an amplitude of up to a few magnitudes in the visible wavelength, and somewhat less in the near-infrared bands \citep{cioni03}, with pulsation amplitudes $\sim~3$ times smaller for RGB than AGB stars (e.g.~\citealp{wood83}). For both carbon-rich and oxygen-rich AGB stars, it is thought that an interplay between pulsation and convection leads to strongly extended molecular atmospheres with temperatures cool enough to form dust. The radiation pressure on dust grains is often believed to drive the mass loss (see, e.g., \citealp{hoefner11, hoefnerandolofsson18, bladh19} and references therein, for discussion of the success and the current difficulties with this scenario). However, observational evidence from interferometric observations and dynamical modeling the atmospheres (e.g.~\citealp{sacuto11, rau15, wittkowski17, rau17, wittkowski18}), have raised issues with this premise. 
For RSGs, it has been speculated that the same processes may explain their mass loss as well. However, \cite{arroyo-torres15, ohnaka17} showed that 1D and 3D dynamic model atmospheres of RSGs based only on pulsation and convection cannot explain both the observed radial extensions of RSG atmospheres or how the atmosphere is extended to radii where dust can form (see Fig.~\ref{COBOLD_2}).

\hspace{\parindent} These challenges point towards missing observational and theoretical considerations in our understanding of cool evolved stars.

\section{Pathway Towards Understanding}
 Heretofore overlooked processes like radiation pressure on molecular lines \citep{josselin08}, energy transport by Alfv\'{e}n waves \citep{airapetian00, airapetian10}, and magnetic activity \citep{vlemmings18} in cool evolved stars are clear pathways towards unraveling the mysteries held in the atmospheres of cool evolved stars. 

\hspace{\parindent} The importance of magnetic activity and molecular emission is gleaned from observations from the molecule-rich circumstellar shells of AGB and RSG stars.  Evidence for large-scale magnetic fields threading the circumstellar shell comes from spectropolarimetric measurements of the magnetic field at various radii in a number of stars \citep{vlemmings14} and the first detection of a surface magnetic field at the AGB star $\chi$~Cyg \citep{2014A&A...561A..85L}.  
\hspace{\parindent} Several empirical studies of the UV spectral lines from K- and M-giants and supergiants indicate the roles that chromospheres could play in initiating the mass loss in these stars (e.g.,~\citealp{carpenter88a, carpenter95, linsky17}). Moreover, \cite{martinez11} showed that the chromospheric emission from giants is consistent with basal heating by acoustic waves that might represent an essential component of the mass loss mechanism. And the recent discovery of unexpected and ubiquitous UV emission from AGB stars suggests similar processes could exist in these stars \citep{montez17} and could play a role in the physical properties of the circumstellar environment \citep{2019ApJ...873...36V}. Chromospheric-like emission has been detected and partially-resolved in high angular resolution interferometric ALMA observations of the AGB star W Hya \citep{2017NatAs...1..848V}, a tantalizing result that merits multiwavelength followup (see Sect.~\ref{obs}).  

\subsection{Observational Challenges}\label{obs}

\hspace{\parindent} The recent advances of cool evolved stars studies largely come from piecemeal approaches targeting well-studied bright targets. This approach is partially governed by the sensitivity of existing groundbreaking instrumentation, which favors the brightest targets in order to make the difficult measurements -- often for the first time -- on these stars. In the next decade, as interferometric and multi-wavelength spectroscopic capacity grows, large complete samples of RGB, AGB, and RSG stars will be sought in order to fully characterize the energy transport in the outflows from these stars.  These systematic surveys should cover a large variety of parameter space in terms of bolometric luminosities, spectral types, dust composition, and dust production rates.

\hspace{\parindent} UV spectroscopic observations of chromospheric emission lines, as those done for RGB stars with the HST/GHRS (Goddard High Resolution Spectrograph) and HST/STIS (Space Telescope Imaging Spectrograph) instruments (e.g.~\citealp{carpenter18, rau18}), are essential for
determining how the velocity profile varies with height and the wind acceleration. These details are crucial for understanding where and how energy is imparted into the outflow. After the outstanding legacy and continuous performance of the HST, the proposed LUVOIR (Large UV Optical Infrared telescope) mission will be vitally important in the next decade to continue such UV studies to a wider range of stars.

\hspace{\parindent} The structure, geometry, and density distribution of the circumstellar envelopes are crucial constraints on the mass-loss process in these cool evolved stars. Measuring these properties at different angular scales and evolutionary stages will enable investigations of the mass-loss process in unprecedented detail from deep within the star to the interface with the ISM. 
\hspace{\parindent} The groundbreaking combination of UV spectroscopy and stellar interferometry becomes a powerful tool for investigating stellar chromospheres, winds, and their interface with the photosphere and a circumstellar shell. The vastly improved capacity of recent built and future instruments will be essential for studies of cool evolved stars in our galaxy. For instance, instruments like CHARA/VEGA \citep{charavega} could image stars in the visible Ca~II triplet or H~$\alpha$ lines to understand the angular extension of chromospheres (see eg.~\citealp{berio11}). Also, in the $H$-band ($1.65~\mu$m) with the 6-beam combiner MIRC-X \citep{MIRC, MIRCX}, could image stellar photospheres. In the next decade, extraordinary stellar image details can be produced by interferometers such as the Magdalena Ridge Observatory Interferometer (MROI, \citealp{MROI}), a 10-element imaging interferometer to operate between $0.6$ and $2.4~\mu$m; and upgrades on the Navy Precision Optical Interferometer (NPOI, \citealp{NPOI}) will bring essential, new angular resolution capabilities. Moreover, the Very Large Telescope Interferometer (VLTI) now offers dramatic improvement in optical/infrared interferometry, with its second generation instruments. In particular, the four-telescope beam combining instrument MATISSE \citep{lopez14}, due to its broad wavelength coverage in the thermal infrared ($3$--$13~\mu$m), will be 
able to produce images, for the first time in the thermal infrared, with angular resolution of $\sim3$~mas at $L$-band, having over $10$~pixels across the photosphere of the larger AGB stars, allowing for complex model-independent image reconstructions. 
\hspace{\parindent} In addition, ALMA has demonstrated its enormous potential for resolving the region close to the star (e.g.~\citealp{kervella18}) and the interface between the atmospheric structure of these stars and the ISM where the shaping mechanisms operate (e.g.~\citealp{maercker12, brunner19}). Further studies with ALMA will determine the importance of surface rotation \citep{vlemmings18, ramstedt18} and large-scale circumstellar magnetic fields (e.g.~\citealp{vlemmings12}).

\hspace{\parindent} In the next decade, and beyond, the James Webb Space Telescope (JWST), the Wide
Field Infrared Survey Telescope (WFIRST), and the Extremely Large
Telescope (ELT) will dramatically revolutionize our understanding of RSG and AGB stars in nearby galaxies (e.g.~\citealp{boyer15, dellagli18}) by increasing the sample of these stars at low and high metallicity and by reaching faint targets well beyond the Local Group.

\subsection{Theoretical Challenge} 
\hspace{\parindent} The aforementioned processes have yet to be incorporated into theoretical models such as the 1D hydrostatic PHOENIX models, 3D convection models (CO5BOLD, \citealp{freytag12}), 1D self-excited pulsation models of RSGs (CODEX model series by \citealp{ireland08, ireland11}), and 1D DARWIN models \citep{hoefner16}. Presently, multi-dimensional, radiation-hydrodynamics (RHD) codes like CO5BOLD have successfully modeled the outer convective envelope and dust-forming atmosphere of an M-type AGB star \citep{hoefnerandfreytag19}, suggesting that convection and pulsations are the likely mechanism for producing the observed clumpy dust clouds formed by large-scale, non-spherical shock waves that generate grain growth in their wakes. However, the empirical evidence of the full range of other physical and dynamical processes outlined in the previous sections requires their incorporation into 3-D dynamic model atmospheres to test such models against time series of interferometric observations that spatially resolve stellar disks. 

\hspace{\parindent} Another missing element in models of stellar evolution is the aftermath of the AGB and RSG stages. For AGB stars, their asymmetric mass loss should produce the rich variety of PNe shapes \citep{balickfrank}, while the mass loss asymmetries in RSG \citep{humphreys07} may influence their subsequent core-collapse supernova explosions (e.g.~\citealp{walmswell12, morozova17}). In both of these processes, the potential effect of binary companions, rotation, and magnetic activity can be investigated theoretically \citep{demarco, ramstedt17}. High spatial resolution interferometric observations are required to resolve such systems, and compare observations with theorical models of binary stellar evolution.


\section{Recommendations}
\hspace{\parindent} We make the following recommendations to the Astro2020 Decadal Survey Committee:

\textbf{(I)} Support theoretical and observational studies to: (a) identify and model the physical processes that need to be included in current 1D and 3D model atmospheres of cool, evolved stars; and (b) characterize the shaping mechanisms for PNe and their relation with the parent AGB stars, and the progenitors in the environment of core-collapse and Type~Ia SNe.

\textbf{(II)} Support observational programs that test and develop our understanding of the properties and parameters of photospheric, chromospheric, and outer atmospheric regions of evolved cool stars through high-resolution spectroscopic observations and interferometric observations. 

\textbf{(III)} Support theoretical modeling to include the neglected role of chromospheric activity in the mass loss process of RGB and RSG stars, and to understand the roles of their magnetic fields. 

\textbf{(IV)} Support for programs that include coordinated multi-messenger ground- and space-based observations. 

\pagebreak
\bibliographystyle{aasjournal}
\bibliography{biblio}

\end{document}